\documentclass[10pt]{article}

\usepackage[preprint, nonatbib]{neurips_2025}

\usepackage[utf8]{inputenc} 
\usepackage[T1]{fontenc}    
\usepackage{hyperref}       
\usepackage{url}            
\usepackage{booktabs}       
\usepackage{amsfonts}       
\usepackage{nicefrac}       
\usepackage{microtype}      
\usepackage{xcolor}         
\usepackage{todonotes}         
\usepackage{amsmath}
\usepackage{subcaption}
\usepackage[style=numeric-comp,sorting=none]{biblatex}
\begin{document}

\title{Rapid Inference of Logic Gate Neural Networks for Anomaly Detection in High Energy Physics}

\author{%
  Lino Gerlach \\
  Princeton University\\
  \texttt{lg0508@princeton.edu} \\
  \And
  Elliott Kauffman \\
  Princeton University \\
  \texttt{ek8842@princeton.edu} \\
  \And
  Liv V\r{a}ge \\
  Princeton University \\
  \texttt{lv7805@princeton.edu} \\
  \And
  Isobel Ojalvo \\
  Princeton University \\
  \texttt{iojalvo@princeton.edu} \\
}

\date{\today}

\maketitle

\begin{abstract}
The increasing data rates and complexity of detectors at the Large Hadron Collider (LHC) necessitate fast and efficient machine learning models, particularly for rapid selection of what data to store, known as triggering. Building on recent work in differentiable logic gates, we present a public implementation of a Convolutional Differentiable Logic Gate Neural Network (CLGN). We apply this to detecting anomalies at the Level-1 Trigger at CMS using public data from the CICADA project. We demonstrate that the CLGN achieves physics performance on par with or superior to conventional quantized neural networks. We also synthesize an LGN for a Field-Programmable Gate Array (FPGA) and show highly promising FPGA characteristics, notably zero Digital Signal Processor (DSP) resource usage. This work highlights the potential of logic gate networks for high-speed, on-detector inference in High Energy Physics and beyond.
\end{abstract}

\section{Introduction}

The upgrade of the Large Hadron Collider (LHC) \cite{lhc-OG} to the High-Luminosity LHC (HL-LHC) \cite{hl-lhc-tdr} will present large data processing challenges. The CMS Level-1 (L1) Trigger \cite{l1t}, which is the first level of data reduction at the CMS experiment, will need to process 63 Tb of data per second \cite{tdr-p2-l1t}. Machine learning (ML) is often a great candidate for processing large data quantities, but making the inference latency low enough for triggering applications without losing the required physics performance is a challenge. This has lead to a series of innovations for ML on FPGAs, such as the high level synthesis package for ML called hls4ml \cite{hls4ml}. 

One very promising low latency ML infrastructure is Differentiable Logic Gate Neural Networks (LGNs), first introduced in \cite{lgn}. Unlike traditional neural networks, LGNs use logic gates as their fundamental computational units. During training, these gates are relaxed to continuous, differentiable approximations, but at inference, they become purely binary. This avoids expensive floating-point operations and has no computationally expensive matrix multiplications. It results in rapid inference, and the architecture has achieved SOTA inference times on known problems, such as MNIST digit classification. The framework was later extended to Convolutional Differentiable Logic Gate Networks (CLGN) in \cite{clgn}.

In this work, we present our re-implementation of the CLGN architecture, motivated by the lack of publicly available CLGN code. We then apply this model to the Anomaly Detection (AD) task from the CMS Calorimeter Image Convolutional Anomaly Detection Algorithm (CICADA) project \cite{cicada}. The CICADA project uses a "teacher-student" model setup, where a large autoencoder - the "teacher" -  learns to reconstruct calorimeter images, and a small, low-latency model - the "student" - is trained to emulate its output for on-detector inference. We utilize our CLGN as the student model and report on its physics performance as well as demonstrate the performance of an LGN student model on FPGA. 

\section{Differentiable Logic Gate Networks}

LGNs replace traditional neurons with logic gates. Each node in the network receives two inputs and learns to approximate one of 16 possible two-input Boolean logic gates. In order to learn, the system must be differentiable, and logic gates are not. This is solved by relaxing the binary logic to a continuous approximation during training. An overview of the logic gates and their approximations can be found in \cite{lgn}. The output of a logic gate block, $a^{\prime}$, is a convex combination of the 16 possible gates, with each gate's contribution weighted by learnable parameters $w_i$. This is typically represented as a softmax over the gates' real-valued functions:
\begin{equation}
a^{\prime}=\sum_{i=0}^{15}\frac{e^{w_{i}}}{\sum_{j=0}^{15}e^{w_{j}}}\cdot f_{i}(a_{1},a_{2})
\end{equation}
where $a_1$ and $a_2$ are the real-valued inputs and $f_i(a_1,a_2)$ is the continuous, real-valued function of the i-th logic gate. For example, a relaxed AND gate is the product $a_1 \cdot a_2$, while a relaxed OR is $a_1 + a_2 - a_1 \cdot a_2$. This differentiable formulation allows for standard gradient-based optimization using techniques like backpropagation.

During inference, the network operates in a purely binary mode. The softmax is replaced by a hard argmax operation of the learned weights. Each node selects only the single most probable logic gate from the 16 possibilities, and this gate operates on binary inputs. This transformation from continuous to binary logic gates at inference time enables ultra-low-latency, bit-level computation.

\subsection{Convolutional Logic Gates and binarization}\label{subsec: clgn and binarization}

To extend the LGN concept to high-dimensional, image-like data, such as the calorimeter images from CMS, we implement Convolutional Differentiable Logic Gate Networks (CLGNs) as described in \cite{clgn}. A standard convolutional kernel, which typically performs a matrix multiplication, is replaced with a binary tree structure of logic gates that aggregates information across a receptive field.

A crucial prerequisite for these networks is the conversion of continuous input data into a binary representation. We employ thermometer encoding for this purpose, a technique previously shown to be effective in quantized neural networks \cite{thermalEncoding}. Instead of binarizing by set thresholds, a thermometer encoding learns what the thresholds should be. For an input value x and a set of N learnable thresholds $T=\{t_1,t_2, \ldots ,t_N\}$, the thermometer encoding is a binary vector of length N, where the i-th bit is 1 if $x \ge t_i$ and 0 otherwise. For instance, with thresholds $T=\{10,20,30\}$, an input value of 15 would be encoded as 100, while a value of 25 would be 110. 

\section{Application to the CICADA Project}

The CICADA project provides anomaly detection in the CMS L1 Trigger using a teacher-student model architecture. The teacher model is a large autoencoder that reconstructs $18 \times 14$ pixel calorimeter images and calculates an anomaly score based on the reconstruction loss. The student model is a low-latency convolutional neural net deployed on an FPGA, and is trained to replicate the teacher's anomaly score. We use our re-implemented CLGN as the student model. The teacher and student models are trained on the 2017 CMS Open data \cite{CERNOpenData2017}. The teacher model is trained on zero bias proton-proton events, which serve as the inlier events. The student is trained on this as well as outliers, which here are top quark pair production, $t\bar{t}$, events. $t\bar{t}$ has three possible final states characterized by the number of leptons (l), neutrinos ($\nu$), and quarks (q). A $t\bar{t} \rightarrow 2l + 2\nu$ sample is used in training, validation and testing of the student model, while $t\bar{t} \rightarrow l + \nu + 2q$ and $t\bar{t} \rightarrow 4q$ are used only in the test set. 

\begin{figure}[ht!]
\centering

\begin{subfigure}[t]{0.48\textwidth}
  \centering
  \vspace{0pt} 
  \includegraphics[width=\textwidth]{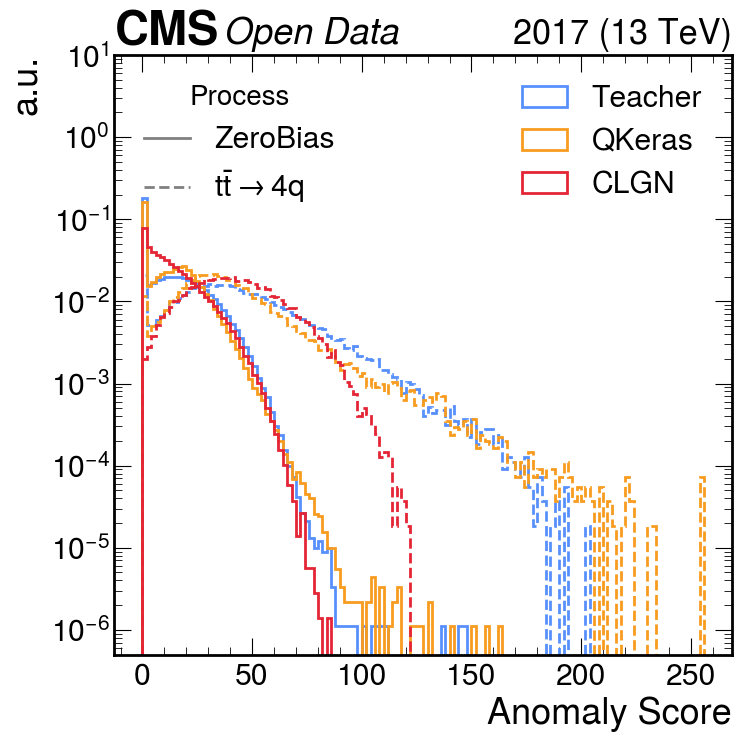}
  \caption{Anomaly scores of outlier and inlier signals.}
  \label{fig:anomaly_score}
\end{subfigure}%
\hfill
\begin{subfigure}[t]{0.48\textwidth}
  \centering
  \vspace{0pt}
  \includegraphics[width=\textwidth]{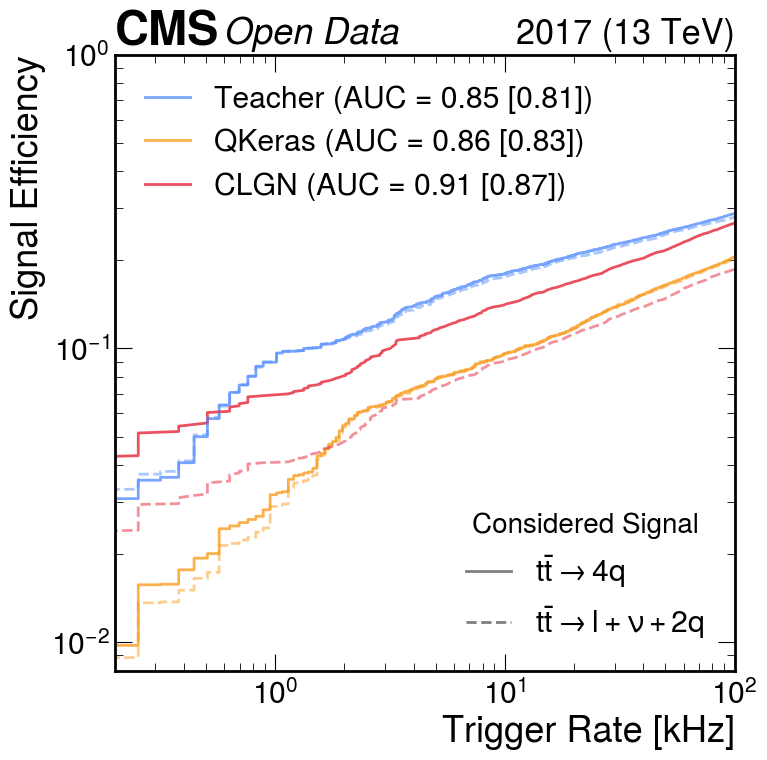}
  \caption{Signal efficiency of two outlier samples. The two AUC scores refer to the score on each sample, with the dashed line AUC in square brackets.}
  \label{fig:trigger_rate}
\end{subfigure}

\vspace{1.2em} 
\caption{Performance of a CLGN compared to the teacher model and a quantized non-LGN implementation.}
\label{fig:cicada-results}
\end{figure}

The results are shown in \autoref{fig:cicada-results}. The model is compared to the current CICADA model implemented in the trigger, which is a quantized model \cite{qkeras}. More details on this can be found in \cite{cicada} and \cite{linoNeurips}. \autoref{fig:anomaly_score} shows that the anomaly score for the inlier data is fairly similar across models. The anomaly score for the outliers is generally lower for the CLGN, which is related to how the data is binarized. A good model has good separation between the inlier and outlier anomaly scores, which all these models have. This is illustrated further in \autoref{fig:trigger_rate}. The signal efficiency is the rate at which one correctly triggers on a signal. The triggering rate is the frequency with which one triggers, as determined by the anomaly score threshold one decides to trigger on. \autoref{fig:trigger_rate} is therefore analogous to a ROC curve, with true positive and false positive rates, and the area under the curve corresponds to model quality. The CLGN model outperforms the QKeras model \cite{qkeras, linoNeurips} for both samples, especially at low trigger rates. 

\section{FPGA Implementation and Performance}

Satisfied that the physics performance of the CLGN was good, work was started to synthesise the model for FPGA. Since the CLGN model is quite complicated, this work is still ongoing. The work shown here instead uses an LGN, which has about half the number of trainable parameters as the model shown in the previous section. Details can be found in the code implementation described in \autoref{app:code}. 

We synthesized our CLGN model for an AMD Virtex-7 FPGA using AMD Vitis HLS \cite{XilinxVitisHLS2021}. AMD Vitis HLS is a tool that translates C or C++ code to a hardware description language (HDL) for deployment on FPGAs. This process, known as High-Level Synthesis (HLS), automates the complex task of designing custom hardware circuits. We therefore first translated our code to C, then synthesized for FPGA. 

\begin{figure}
    \centering
    \includegraphics[width=0.9\linewidth]{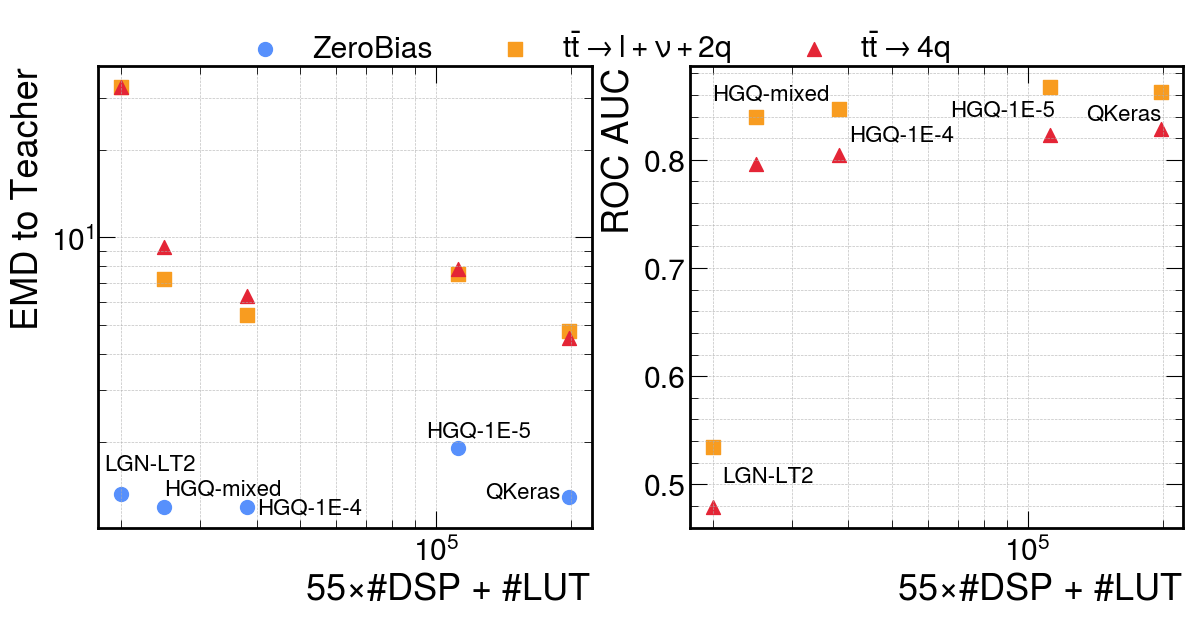}
    \caption{The FPGA resource utilisation of our LGN implementation compared to a quantized network (QKeras) and a High Granularity Quantized implementation. HGQ-1E-4, HGQ1E-5 and HGQ-mixed refer to a specific hyperparameter in the granularization process, see \cite{linoNeurips} for more details. EDM denotes the earth mover distance. The resource utilization on the x-axis, while somewhat arbitrary, is a common metric for FPGA utilization.}
    \label{fig:resource util}
\end{figure}

A major advantage of the LGN architecture is the complete absence of floating-point arithmetic and matrix multiplications at inference time. These operations are typically executed on specialized FPGA resources called Digital Signal Processors (DSPs), which are both a limited and expensive resource on the chip. Our CLGN implementation achieves zero DSP usage, and is instead implemented entirely using Look-Up Tables (LUTs) and Flip-Flops (FFs). LUTs are general-purpose combinatorial logic elements that can implement any Boolean function, while FFs are used for storing states. This is shown in \autoref{fig:resource util} and \autoref{tab:fpga_reousrce}. \autoref{fig:resource util} shows that the LGN model achieves very similar performance compared with the teacher model as the other models, but with far fewer resources. It performs worse on the outliers, likely because it is a feedforward network instead of a convolutional one. \autoref{tab:fpga_reousrce} shows that the latency of the LGN model is very low compared to the alternatives. 

\begin{table}[ht!]
\centering
\caption{Vitis HLS re-place-and-route hardware cost comparison for select student models. Latency is reported
in clock cycles (cc), where one cycle is 6.25 ns}
\label{tab:fpga_reousrce}
\begin{tabular}{l  ccccc}
    Model Label & Quantization Library & Latency (in cc) & DSPs  & FFs & LUTs \\
    \hline
    QKeras          & QKeras               & 16 & 697 & 50368 & 159447 \\
    HGQ-1E-5        & HGQ                  & 17 & 4 & 27776 & 111848 \\
    HGQ-1E-4        & HGQ                  & 11 & 1 & 6229 & 38111 \\
    HGQ-mixed       & HGQ                  & 8 & 0 & 3019 & 24947 \\
    LGN-LT2         & LGN                  & 3 & 0 & 856 & 19977 \\
\end{tabular}
\end{table}

\section{Conclusion}

We have successfully re-implemented and deployed a Convolutional Differentiable Logic Gate Network for the CICADA anomaly detection task. Our results demonstrate that this architecture not only achieves excellent physics performance but also offers substantial advantages in terms of on-detector inference, including ultra-low latency and zero DSP usage on an FPGA. This work confirms the viability of logic gate neural networks for high-speed applications in HEP and paves the way for further exploration of these resource-efficient models.

\section*{Acknowledgments}
This work was supported by the National Science Foundation under Cooperative Agreements OAC-1836650 and PHY-2323298.

\printbibliography
\appendix
\section{Code implementation}\label{app:code}
The original implementation of LGNs can be found in a github repository called \href{https://github.com/Felix-Petersen/difflogic}{difflogic}. This does not contain the implementation of CLGNs described in \cite{clgn}. We therefore replicated this work, and the complete work can be found in our repository called \href{https://github.com/ligerlac/torchlogix}{torchlogix}.This is under continous development. 









\end{document}